\documentclass[5p,number]{elsarticle}

\usepackage{hyperref}% add hypertext capabilities
\usepackage[T1]{fontenc}
\usepackage[utf8]{inputenc}
\usepackage{textcomp} % for \textdegree
\usepackage{lmodern} % nicer font - The Latin Modern fonts are enhanced versions of the Computer Modern fonts. They have enhanced metrics and glyph coverage.
\usepackage{amsmath}
\usepackage{amsfonts}
\usepackage{breakurl} %  for breaking urls
\usepackage[switch]{lineno} %lines numbering, switch makes right numbering on right columns
\graphicspath{{./figures/}}
\biboptions{sort&compress}

\hypersetup{
    breaklinks = true,
    colorlinks = true,
    pdftitle = {}, 
    pdfauthor = {},
    pdfkeywords = {},
    linkcolor = blue,
    citecolor = blue,
    filecolor = black,
    urlcolor = magenta
}

\makeatletter % reset of second corresponding author
\def\@author#1{\g@addto@macro\elsauthors{\normalsize%
    \def\baselinestretch{1}%
    \upshape\authorsep#1\unskip\textsuperscript{%
      \ifx\@fnmark\@empty\else\unskip\sep\@fnmark\let\sep=,\fi
      \ifx\@corref\@empty\else\unskip\sep\@corref\let\sep=,\fi
      }%
    \def\authorsep{\unskip,\space}%
    \global\let\@fnmark\@empty
    \global\let\@corref\@empty  %% Added
    \global\let\sep\@empty}%
    \@eadauthor={#1}
}
\makeatother

\newcommand{\lani}{La$_2$Ni$_7$} 
\newcommand{\lamgni}{La$_{2-x}$Mg$_x$Ni$_7$} 
\newcommand{\lamghalf}{La$_{1.5}$Mg$_{0.5}$Ni$_7$}
\newcommand{\lagdmgni}{La$_{1.5-x}$Gd$_x$Mg$_{0.5}$Ni$_7$} 
\newcommand{\lagdquartermgni}{La$_{1.25}$Gd$_{0.25}$Mg$_{0.5}$Ni$_7$}
\newcommand{\lamghalfnico}{La$_{1.5}$Mg$_{0.5}$Ni$_{7-y}$Co$_y$}
\newcommand{\lamghalfnicohalf}{La$_{1.5}$Mg$_{0.5}$Ni$_{6.5}$Co$_{0.5}$}
\newcommand{\lacommamgni}{(La,Mg)$_2$Ni$_7$}
\newcommand{\gdco}{Gd$_2$Co$_7$} 
\newcommand{\ceni}{Ce$_2$Ni$_7$} 
\newcommand{\ab}{A$_2$B$_7$}

\begin{document}
\begin{sloppypar} %for overfull lines with long chemical formulas

\title{Effect of Gd and Co content on electrochemical and electronic properties of \lamghalf{} alloys: a combined experimental and first-principles study}

\author{Miros\l{}aw Werwi\'nski\corref{cor1}}
\cortext[cor1]{Corresponding author} 
\ead{werwinski@ifmpan.poznan.pl}

\author{Andrzej Szajek}
\author{Agnieszka Marczyńska}
\author{Lesław Smardz}

\address{Institute of Molecular Physics, Polish Academy of Sciences,\\  M. Smoluchowskiego 17, 60-179 Pozna\'n, Poland}

\author{Marek Nowak}
\author{Mieczysław Jurczyk}

\address{Institute of Materials Science and Engineering, Poznań University of Technology,\\
Jana Paw\l{}a II No. 24, 61-138 Poznań, Poland}

\begin{abstract}

In this work we investigate the effect of Gd and Co substitutions on the electrochemical and electronic properties of \lamghalf{} alloy. 
Two series of \lagdmgni{} ($x$~=~0.0, 0.25, 1.0) and \lamghalfnico{} ($y$~=~0.0, 0.5, 1.5) alloys are produced using mechanical alloying technique. 
The X-ray diffraction indicates multi-phase character of the samples with majority of hexagonal \ceni{}-type and rhombohedral \gdco{}-type structures of \lacommamgni{} phase. 
Partial substitutions of La by Gd or Ni by Co in \lamghalf{} phase result in increase of cycle stability of the metal hydride (MH$_x$) electrodes. 
All considered alloys reach the maximum discharge capacity after three charging-discharging cycles. 
Two optimal compositions, in respect of electrochemical properties, are subsequently investigated by the X-ray photoelectron spectroscopy (XPS).
The experimental analysis of the valence band is further extended by the density functional theory (DFT) calculations.
We also discuss the effects of alloying and site preference of dopants on the position of the van Hove-type singularity as observed in the electronic densities of states (DOS) in proximity of the Fermi level.

\end{abstract}

\begin{keyword} 
Metal hydrides \sep Mechanical alloying  \sep XPS  \sep DFT  \sep La-Mg-Ni alloys
\end{keyword}

\date{\today}

\maketitle

\section{Introduction}

The next generation of electrode materials for nickel-metal hydride (Ni-MH$_x$) batteries are the \lacommamgni{}-type hydrogen storage alloys~\cite{ouyang_progress_2017, zhang_new_2017,fan_phase_2017, liu_phase_2016}.
It has been found that the limited cycle stability of the AB$_2$-type and AB$_5$-type materials could be compensated in the \ab{}-type alloys~\cite{liu_phase_2016}.
It has been shown that substitution of La with other rare-earth element~\cite{liu_phase_2016, xue_phase_2017, lv_influence_2018, liu_effect_2015, wu_electrochemical_2017, li_effect_2008} 
and Ni with alternative 3$d$ element~\cite{guzik_effect_2017,li_effect_2017, lv_microstructural_2017} also modifies the electrochemical performance of the \ab{} materials. 
However, a cycling stability and high-rate discharge capacity, especially for high discharging rates and material and processing costs, still need improvements to satisfy increasing demands of the market~\cite{cao_enhanced_2015, huang_graphene/silver_2017}.

The most common phases of the multi-phase La-Mg-Ni-based alloys are (La,Mg)Ni$_3$, \lacommamgni{} and (La,Mg)$_5$Ni$_{19}$~\cite{ma_synergic_2017,li_effect_2016}.
They are all composed of [A$_2$B$_4$] and [AB$_5$] subunits stacked alternatively along the $c$~axis and forming superlattice structures~\cite{yartys_structureproperties_2015}.
Studies on the electrochemical performance of the \ab{}-type single-phase \lamghalf{} alloy have revealed that both the (La,Mg)$_5$Ni$_{19}$ and the LaNi$_5$ phases have catalytic effects on the fast discharge of the alloy~\cite{liu_effect_2015}.
A partial replacement of La by the rare earth (RE) atoms and Ni by Co or Al can affect the crystal structure of the \ab{} alloys.
Nd and Co doped alloys can be applied as negative electrodes in Ni-MH$_x$ batteries, which are characterized by improved cyclic durability~\cite{wu_electrochemical_2017,dong_cooperative_2011}.
By adjusting the RE concentration in the (RE-Mg)$_2$Ni$_7$ alloys it is possible to enhance the electrochemical hydrogen storage properties of the alloy electrodes.
For La$_{0.8}$Gd$_x$Mg$_{0.2}$Ni$_{3.15}$Co$_{0.25}$Al$_{0.1}$ alloys ($x$ from 0.0 to 0.4) it has been shown that Gd substitution significantly affects the phase composition of the alloys~\cite{li_effect_2016}.
Additionally, the partial substitution of Ni by Co atoms in the \ab{} alloy have also resulted in improving the electrochemical discharge capacity~\cite{anik_synthesis_2015}.
An addition of Co atoms to AB$_5$ alloys reduces its volume expansion caused by hydriding,
which in turn reduces the pulverization and corrosion of the alloys~\cite{liao_effect_2004,wang_phase_2006}. 
Lately, it has been shown that the Pr, Nd and Co-free A$_2$B$_7$-type, La$_{13.9}$Sm$_{24.7}$Mg$_{1.5}$Ni$_{58}$Al$_{1.7}$Zr$_{0.14}$Ag$_{0.07}$ alloy exhibits excellent hydrogen-storage and electrochemical performance. 
The high rate discharge abilities were 97\%, 86\% and 78\% at discharge current densities of 330, 970 and 1600~mAh/g, respectively~\cite{ouyang_hydrogen_2018}.
The improvement of hydrogenation properties of the (La,Mg)$_2$Ni$_7$ system may be achieved by encapsulation of material particles with thin amorphous nickel coating~\cite{dymek_encapsulation_2018}. 
The magnetron sputtering of La$_{1.5}$Mg$_{0.5}$Ni$_7$ particles with 0.29~$\mu$m thick Ni film appeared to have been efficient enough to inhibit corrosion degradation.

Recently, also the effect of Mg substitution on the electrochemical and electronic properties of \lamgni{} materials has been investigated~\cite{balcerzak_hydrogenation_2017,werwinski_effect_2018}.
The X-ray diffraction measurements have indicated a multiphase character of the samples with majority of \lacommamgni{} phases in hexagonal \ceni{}-type and rhombohedral \gdco{}-type structures. 
Electrochemical measurements have shown that the maximum discharge capacity (C$_\mathrm{max}$) increases with Mg concentration and reaches the highest value for the Mg concentration $x$ equal to 0.5. 
Furthermore, the valence band of the nanocrystalline \lamghalf{} sample has been investigated by X-ray photoelectron spectroscopy (XPS) and from density functional theory (DFT)~\cite{werwinski_effect_2018}.
The valence band of isostructural hexagonal La$_2$Co$_7$ compound has been investigated already by DFT~\cite{kuzmin_magnetic_2015}.
A systematic DFT study of every ordered configuration of (La,Mg)$_2$Ni$_7$ and several other La-Mg-Ni systems has been presented by Crivello \textit{et al.}~\cite{crivello_structural_2011}.
The authors have concluded that the stability of the (La,Mg)$_2$Ni$_7$ system decreases with Mg substitution.
Crivello \textit{et al.} have also presented a DFT study on the distribution of hydrogen in the La$_2$Ni$_7$, Mg$_2$Ni$_7$ and \lamghalf{} hosts~\cite{crivello_first_2015}.

An effective method of synthesis of the hydrogen storage materials with reduced crystallite sizes is mechanical alloying (MA)~\cite{varin_nanomaterials_2009}.
Recent studies suggest that MA technique can improve the kinetics of hydrogen absorption and desorption of the synthesized alloys.
This change comes from the large specific surface areas and from short hydrogen diffusion distances in the MA samples~\cite{ma_synergic_2017}. 
In this work we apply the MA technique with subsequent annealing in the argon atmosphere to produce the \lagdmgni{}  ($x$~=~0.0, 0.25, 1.0) and \lamghalfnico{} ($y$~=~0.0, 0.5, 1.5) alloys. 
Subsequently, we investigate the influence of the substitutions on electrochemical and electronic properties of \lacommamgni{}-type materials. 
With X-ray photoelectron spectroscopy and DFT calculations we study the valence band of the selected optimal compositions.
In our theoretical analysis we put special emphasis on narrow energy region in the vicinity of the Fermi level.

\section{Experimental and computational details}
%
%------------general overview-------------
%
This section covers details of preparation of \lagdmgni{} and \lamghalfnico{} samples, the basic informations on the applied characterization techniques: X-ray diffraction (XRD), X-ray photoelectron spectroscopy (XPS) and electrochemical measurements and the computational parameters for the density functional theory (DFT) calculations.
\subsection{Materials preparation}
La and Gd powders -- grated from rods (Alfa Aesar, 99.9\%), Mg powder (Alfa Aesar, 45~$\mu$m, 99.8\%), Co powder (Aldrich, 1.6~$\mu$m, 99.8\%) and Ni powder (Aldrich, 5~$\mu$m, 99.99\%) were used to synthesized the \lagdmgni{} ($x$~=~0.0, 0.25, 1.0) and \lamghalfnico{} ($y$~=~0.0, 0.5, 1.5) alloy powders. 
The considered compositions were prepared by 48~h of mechanical alloying in the argon atmosphere and by subsequent annealing. 
The ratio of the weight of hard steel balls to the weight of powder was 4.2:1. 
Gd substituted alloys were annealed at 1123~K 
and 
Co substituted alloys at 973~K 
for 0.5~h in high purity argon atmosphere. 
\subsection{Structural characterization and electrochemical measurements}
The crystallographic structure of samples was studied at room temperature using Panalytical Empyrean XRD with Cu\,K$_{\alpha_1}$ ($\lambda = 1.54056$~\AA{}) radiation. 
A scanning electron microscope (SEM, VEGA 5135 Tescan) with energy dispersive spectrometer (EDS, PTG Prison Avalon) was used to characterize the chemical compositions of the obtained powdered materials. 

The electrochemical studies were performed at room temperature in a tri-electrode open cell. 
Electrodes were produced by cold pressing of a mixture of 0.4~g alloy with 0.04~g carbonyl nickel powders into a pellet of 8~mm diameter. 
The counter and reference electrodes were Ni(OH)$_2$/NiOOH and Hg/HgO, respectively.
As electrolyte was used 6~M~KOH. 
Tested electrodes were charged and discharged at a current of 40~mA\,g$^{-1}$. 
A cut-off voltage was -0.7~V \textit{versus} a Hg/HgO reference electrode. 
An extended description of the electrochemical studies has been presented in our previous work~\cite{balcerzak_hydrogenation_2017}. 
The cycle stability of the materials was evaluated by the capacity retaining rate after the $n^\mathrm{th}$ cycle:  $(C_n/C_{max}) \times 100 \%$, 
where $C_n$ is discharge capacity at $n$th cycle and  $C_{max}$ is discharge capacity maximum.
\subsection{XPS measurements \label{ssec:xps}}
The XPS spectra measurements were performed at room temperature using the SPECS EA 10 PLUS energy spectrometer with Al-K$_\alpha$ radiation of 1486.6~eV.
The step size in the XPS valence band measurements was set to 0.05~eV.
The energy spectra were analyzed with a hemispherical analyzer with full width at half maximum FWHM$_{\mathrm{Mg-K_\alpha}}$ equal to 0.8~eV for Ag 3$d_{5/2}$. 
The calibration was performed based on the procedure given by Baer \textit{et al.}~\cite{baer_monochromatized_1975}.
The Au 4$f_{7/2}$ peak was positioned at 84.0~eV and the Fermi level at 0~eV. 
The SPECS ion gun etching system was used to remove the surface layers ($\sim10$~nm) of the studied bulk nanocrystalline samples.
Etching was performed with a beam of Ar$^+$ ions of energy 3~keV and inclined to the sample surface at 45\textdegree{}.
The measurements have been preceded by the backing procedures of the analysis chamber ($T$~=~440~K) which made it possible to achieve a base vacuum of 4~$\times$~10$^{-11}$~mbar. 
More details of the XPS measurements can be found in our previous work~\cite{skoryna_xps_2016}.
\subsection{Density functional theory calculations \label{ssec:dft_details}}
\begin{figure}[ht]
\begin{center}
\includegraphics[trim = 0 10 0 0, clip, width=\columnwidth]{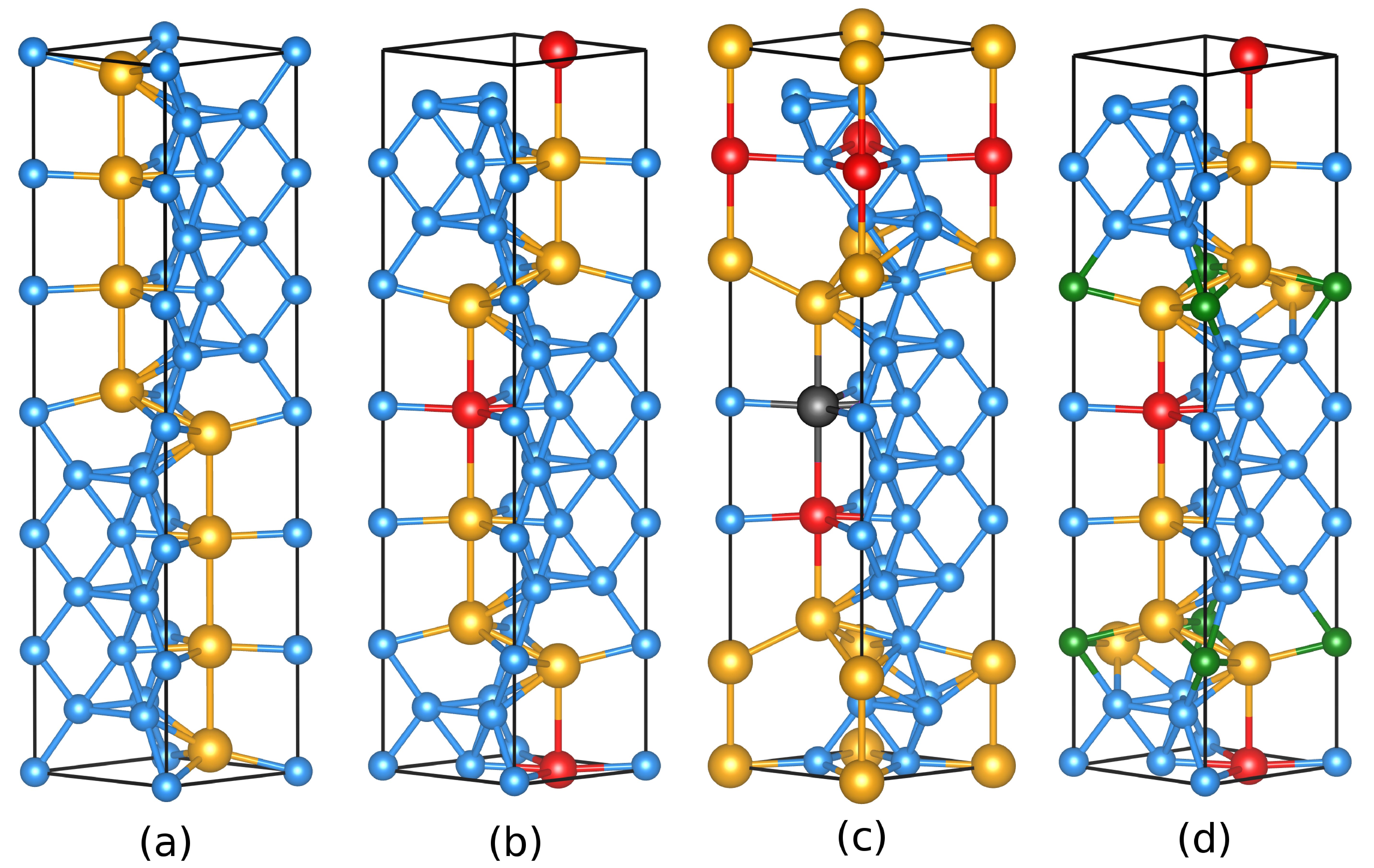} 
\end{center}
\caption{\label{fig:dft_struct}
The crystal structures of
(a) \lani{}; (b) \lamghalf{}; (c) \lagdquartermgni{} and (d) \lamghalfnicohalf{}.
All ordered compound models are considered in Ce$_2$Ni$_7$-type structure.
The different elements are denoted by different colors as follows: La -- yellow, Ni -- blue, Mg -red, Gd -- black and Co -- green.
}
\end{figure}

We performed three sets of calculations with different approaches to model chemical disorder.
The electronic band structure calculations of \lani{}, \lamghalf{}, \lagdquartermgni{} and \lamghalfnicohalf{} compositions were carried out by using the full-potential local-orbital minimum-basis scheme (FPLO18.00-52)~\cite{koepernik_full-potential_1999}. 
These calculations were performed in the scalar relativistic approach within the generalized gradient approximation (GGA) of the exchange-correlation potential in the  form of Perdew-Burke-Ernzerhof (PBE)~\cite{perdew_generalized_1996}.
We used $20 \times 20 \times 5$ \textbf{k}-mesh and energy convergence criterion 10$^{-8}$~Ha ($2.72\times10^{-7}$~eV).
To simulate the chemical disorder we used primarily the so-called ordered compound method.
We started with a model of \lani{} unit cell composed of four formula units (La$_8$Ni$_{28}$).
In order to prepare the compositions corresponding to the experimental ones, within a single unit cell we substituted La by Mg, La by Gd, and Ni by Co atoms.
Figure~\ref{fig:dft_struct} presents the considered models as, visualized using the VESTA computer code~\cite{momma_vesta_2008}.
For simplicity, we considered only the hexagonal Ce$_2$Ni$_7$-type structure and neglected the rhombohedral Gd$_2$Co$_7$-type structure.
%
%-----------------calculated structures--------------------
%
\begin{table}[!ht]
\caption{\label{tab:atomic_pos_ce2ni7} Crystallographic parameters of \lani{} in Ce$_2$Ni$_7$-type structure reproduced after Levin \textit{et al.}~\cite{levin_hydrogen_2004}. 
Lattice parameters $a$~=~5.0577, $c$~=~24.7336~\AA{}, space group \textit{P}6$_3$/\textit{mmc}.\\
}
\centering
\begin{tabular}{rclll}
\hline 
      atom& site &$x$ 		& $y$ 		& $z$ \\
\hline        
        La$_1$& 4$f$ &1/3	& 2/3		& 0.03013 \\
        La$_2$& 4$f$ &1/3	& 2/3		& 0.17422 \\
        Ni$_1$& 2$a$ &0		& 0		& 0 \\
        Ni$_2$& 4$e$ &0		& 0		& 0.16898 \\
        Ni$_3$& 4$f$ &1/3	& 2/3		& 0.83616 \\
        Ni$_4$& 6$h$ &0.8341	& 0.6682	& 1/4 \\
        Ni$_5$& 12$k$ &0.8350	& 0.6700	& 0.08806  \\ 
 \hline      
\end{tabular}
\end{table}
The crystallographic parameters of \lani{} in Ce$_2$Ni$_7$-type structure used for calculations are collected in Table~\ref{tab:atomic_pos_ce2ni7}.
For the ordered compound models with Mg, Gd and Co we used the appropriate lattice parameters from our measurements, see Table~\ref{tab:exp_data}. 
Although the hexagonal \lani{} phase undergoes the antiferromagnetic transition at T$_N$~=~51~K~\cite{parker_magnetic_1983}, we considered only non-magnetic models as consistent with a magnetic state of the samples in room temperature.
However, based on our experience with modeling of Gd alloying~\cite{pierunek_normal_2017}, 
we made an exception for La$_{1.25}$Gd$_{0.25}$Mg$_{0.5}$Ni$_7$ system and simulated it as spin polarized.

%---------------------VCA-----------------------------
%
The second method we used to simulate alloying was the virtual crystal approximation (VCA)~\cite{kudrnovsky_magnetic_2008}.
The idea behind the VCA is to simulate a random homogeneous on-site occupation of two atom types by a single (virtual) atom with an averaged atomic number.
The application of the VCA is, however, limited to the pairs of elements which atomic number differs by one.
In the considered case of La$_{1.5}$Mg$_{0.5}$Ni$_{7-y}$Cu$_y$, in which the Cu/Ni alloying is described by the VCA, 
the atomic number of virtual Cu/Ni atom equals 
$Z_{\text{VCA}} = (1-y) Z_{\text{Ni}} + y Z_{\text{Cu}}$.

%------------------cpa---------------
%
Third and last method used to model chemical disorder was the  coherent potential approximation (CPA)~\cite{soven_coherent-potential_1967}.
The CPA calculations, performed with the FPLO5.00 code~\cite{koepernik_self-consistent_1997, sniadecki_induced_2014}, were intended to find the site preference of Co dopants in the example phase of La$_{1.5}$Mg$_{0.5}$Ni$_{6.5}$Co$_{0.5}$ (Gd$_2$Co$_7$-type).
In contrast to the other presented results, the CPA calculations are performed with the Perdew and Wang form of the exchange-correlation potential~\cite{perdew_accurate_1992} and $6 \times 6 \times 6$ \textbf{k}-mesh (limited by the complexity of the task).

%----------------xps theory------------------
%
The theoretical X-ray photoelectron spectra were obtained from the calculated densities of electronic states (DOS) convoluted by Gaussian with a half-width $\delta$ and scaled using the proper atomic subshell photoionization cross-sections~\cite{yeh_atomic_1985,morkowski_x-ray_2011,skoryna_xps_2016}.
The purpose of the Gaussian convolution was to mimic an experimental resolution, a lifetime of the hole states and the thermal effects.
The value of $\delta$ equal to 0.5~eV was selected to be close to the instrumental value of spectrometer used to take the measurements.

\section{Results and discussion}
In this section we present the results and discussion of structural and electrochemical measurements, XPS valence band characterization and first-principles calculations. 
\subsection{Crystallographic structure and phase composition}
\begin{figure}[!ht]
\begin{center}
\includegraphics[trim = 0 0 0 0, clip,width=0.9\columnwidth]{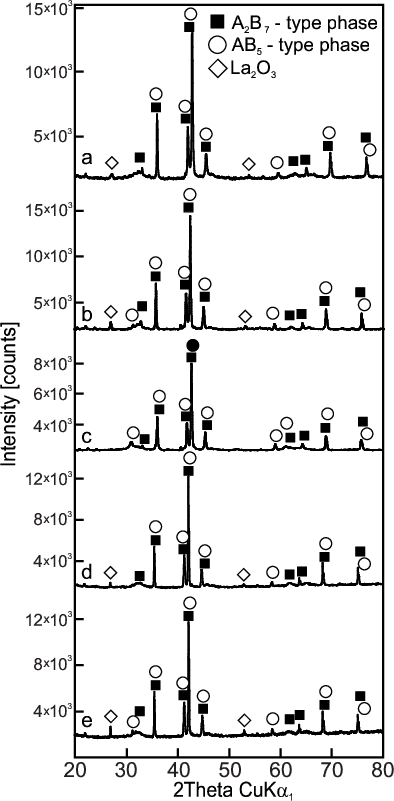} 
\end{center}
\caption{\label{fig:xrd}
The X-ray diffraction patterns of alloys obtained by 48~h of mechanical alloying process and subsequent heat treatment.
(a) \lamghalf{},  
(b) \lagdquartermgni{}, 
(c) La$_{0.5}$Gd$_{1.0}$Mg$_{0.5}$Ni$_7$, 
(d) La$_{1.5}$Mg$_{0.5}$Ni$_{6.5}$Co$_{0.5}$ and 
(e) La$_{1.5}$Mg$_{0.5}$Ni$_{5.5}$Co$_{1.5}$.
}
\end{figure}
\begin{table*}[!ht]
\caption{\label{tab:exp_data} 
Lattice parameters of A$_2$B$_7$ phase, phase abundance, theoretical capacity ($C_\mathrm{th}$), maximum discharge capacity ($C_{max}$), discharge capacity at 30 and 50 cycles ($C_{30}$, $C_{50}$) and $C_{50}/C_{max}$ ratio for the considered alloys.\\
}
\centering
\small
\begin{tabular}{lcccccccccc}
\hline 
\multicolumn{1}{c}{Composition}		&\multicolumn{2}{l}{\footnotesize Lattice parameters} & \multicolumn{3}{l}{\footnotesize Phase abundance (wt\%)}& $C_\mathrm{th}$ & $C_{max}$ 	& $C_{30}$ 	& $C_{50}$ 	& $C_{50}/C_{max}$ \\
						& $a$ (\AA{}) 	& $c$ (\AA{})	&A$_2$B$_7$&AB$_5$&La$_2$O$_3$& (mAh/g)	& (mAh/g)	& (mAh/g)	& (mAh/g)	& (\%)	\\
\hline 			
La$_{1.5}$Mg$_{0.5}$Ni$_7$					&5.04	&24.20	&88.9	&8.0	&3.1		&535		&304		&204		&167		&55\\
La$_{1.25}$Gd$_{0.25}$Mg$_{0.5}$Ni$_7$		&5.01	&24.14	&91.6	&6.4	&2.0		&531		&303		&218		&195		&64\\
La$_{0.5}$Gd$_{1.0}$Mg$_{0.5}$Ni$_7$		&4.97	&24.07	&70.9	&29.1	&0.0		&520		&68			&65			&65			&96\\
La$_{1.5}$Mg$_{0.5}$Ni$_{6.5}$Co$_{0.5}$	&5.05	&24.15	&96.8	&0.8	&2.4		&533		&299		&238		&213		&71\\
La$_{1.5}$Mg$_{0.5}$Ni$_{5.5}$Co$_{1.5}$	&5.04	&24.20	&95.8	&0.9	&3.3		&529		&307		&236		&199		&65\\
\hline 
\end{tabular}
\end{table*}
The \lagdmgni{} ($x$~=~0.0, 0.25, 1.0) and \lamghalfnico{} ($y$~=~0.0, 0.5, 1.5) alloys are produced by the MA and annealing process. 
Figures~\ref{fig:xrd}~(a)--(c) present the XRD spectra of the \lagdmgni{} ($x$~=~0.0, 0.25, 1.0) alloys after 48~h of MA and additional heat treatment at 1123~K for 0.5~h in the argon atmosphere. 
The XRD measurements reveal a multi-phase character of the considered alloys.
The mass fractions of the phases in the La$_{1.5-x}$Gd$_x$Mg$_{0.5}$Ni$_7$ and 
La$_{1.5}$Mg$_{0.5}$Ni$_{7-x}$Co$_x$ MA and annealed alloys were established by the Rietveld refinement of the XRD data in a 2$\theta$ range of 20\textdegree--80\textdegree, see Table~\ref{tab:exp_data}.
The alloys are composed of 
the La$_2$Ni$_7$-type phase 
(\ceni{}-type hexagonal structure (space group \textit{P}6$_3$/\textit{mmc}) 
and the \gdco{}-type rhombohedral structure (s.g. $R$3$m$)), 
the CaCu$_5$-type phase (s.g. $P$6/$mmm$) and 
the La$_2$O$_3$ phase (s.g. \textit{P}$\bar{3}$\textit{m}1). 
The content of the \ab{}-type phase is the highest for $x$~=~0.25 and equal to 91.6~wt\%.
It decreases to 70.9~wt\% for $x=1.0$ and further decreases to 60.0~wt\% for $x=1.5$, see Table~\ref{tab:exp_data} and Ref.~\cite{nowak_hydrogen_2018}.
It has been shown that the La-Mg-Ni based alloys produced by inductive melting or arc melting have similar phase compositions~\cite{xiangqian_structure_2009,iwase_hydrogenation_2012}.

Figures~\ref{fig:xrd}~(d) and (e) show the XRD patterns of \lamghalfnico{} alloys ($y$~=~0.5, 1.5) after 48~h of MA and additional heat treatment at 973~K for 0.5~h in argon atmosphere. 
All of the synthesized alloys are composed of the main \lani{} phase (\ceni{}-type hexagonal structure (s.g. \textit{P}6$_3$/\textit{mmc}) and the \gdco{}-type rhombohedral structure (s.g. $R$3$m$)) and a minor CaCu$_5$-type phase (s.g. \textit{P}6/\textit{mmm}). 
Moreover, a small content of La$_2$O$_3$ phase is observed, see Table~\ref{tab:exp_data}. 
The abundance of the \ab{}-type phase increases from 88.9~wt\%  for \lamghalf{} alloy to 96.8~wt\% for La$_{1.5}$Mg$_{0.5}$Ni$_{6.5}$Co$_{0.5}$. 
The substitution of Co for Ni efficiently prevents the formation of the LaNi$_5$-type phase. 
Previously, a similar tendency has been observed for other Co-substituted \lani{}-type alloys~\cite{anik_synthesis_2015,liao_effect_2004,wang_phase_2006,ming_influence_2011}.
However, it has been shown that a coexistence of the \lani{} phase with the LaNi$_5$ phase increases the electrode stability~\cite{zhang_electrochemical_2015}.

\subsection{Electrochemical measurements}
\begin{figure}[ht]
\begin{center}
\includegraphics[trim = 0 0 0 0, clip,width=\columnwidth]{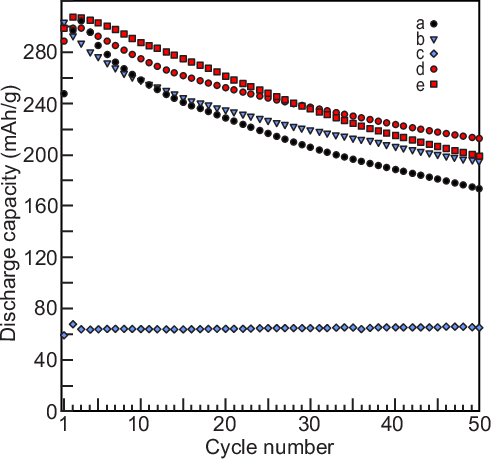} 
\end{center}
\caption{\label{fig:discharge}
Discharge capacities of the alloys as a function of cycle number for
(a) \lamghalf{},  
(b) \lagdquartermgni{}, 
(c) La$_{0.5}$Gd$_{1.0}$Mg$_{0.5}$Ni$_7$, 
(d) La$_{1.5}$Mg$_{0.5}$Ni$_{6.5}$Co$_{0.5}$ and 
(e) La$_{1.5}$Mg$_{0.5}$Ni$_{5.5}$Co$_{1.5}$.
}
\end{figure}

\begin{figure}[ht]
\begin{center}
\includegraphics[width=0.95\columnwidth]{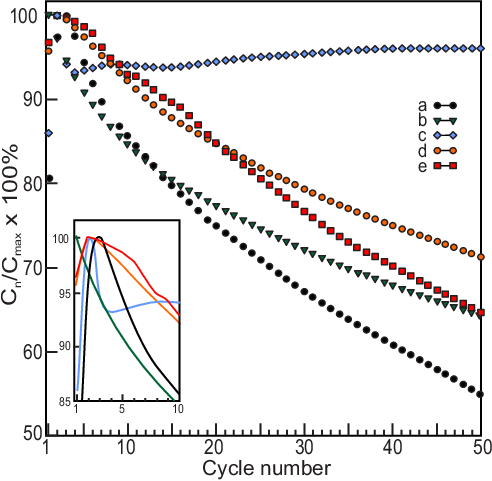} 
\end{center}
\caption{\label{fig:stability}
Cyclic stability of electrodes prepared with:
(a) \lamghalf{},  
(b) \lagdquartermgni{}, 
(c) La$_{0.5}$Gd$_{1.0}$Mg$_{0.5}$Ni$_7$, 
(d) La$_{1.5}$Mg$_{0.5}$Ni$_{6.5}$Co$_{0.5}$ and 
(e) La$_{1.5}$Mg$_{0.5}$Ni$_{5.5}$Co$_{1.5}$.
}
\end{figure}

The results of electrochemical measurements of the \lagdmgni{} ($x$~=~0.0, 0.25, 1.0) and \lamghalfnico{} ($y$~=~0.0, 0.5, 1.5) alloys are summarized in Figs.~\ref{fig:discharge} and \ref{fig:stability} and Table~\ref{tab:exp_data}.
All the considered electrodes display the maximum capacity ($C_{max}$) at the 3rd cycle.
The highest observed values of discharging capacity ($C_{max}$) are 
307~mAh/g for La$_{1.5}$Mg$_{0.5}$Ni$_{5.5}$Co$_{1.5}$ and 
303~mAh/g for \lagdquartermgni{} alloy. 
Previously, for La$_{1.5}$Mg$_{0.5}$Ni$_{7-x}$Co$_{x}$ alloy electrodes ($x$ ranging from 0 to 1.8) the much higher discharge capacities have been measured, in a range from 390 to 406~mAh/g~\cite{zhang_effect_2006}.
For the considered \ab{}-type alloys we evaluate a theoretical capacity ($C_\mathrm{th}$) in the same way as it has been done previously for LaNi$_5$~\cite{willems_metal_1984}.
For \lani{} we use the maximum hydrogen capacity 1.4 H/M (hydrogen-to-metal atom ratio) as measured by Iwase~\textit{et al.}~\cite{iwase_situ_2013}.
1.4 H/M leads to a number of hydrogen atoms per formula unit equal to 12.6 (La$_{2}$Ni$_{7}$H$_{12.6}$).
The same maximum hydrogen capacity we assume for other considered \ab{}-type alloys.
An obtained theoretical capacity for pristine \lani{} is equal to 490~mAh/g and for all considered compositions with Mg, Gd, and Co the $C_\mathrm{th}$ is of about 530~mAh/g, see Table~\ref{tab:exp_data}.
The measured discharge capacity of all studied electrodes is degrading during the charge-discharge cycles.
The mechanisms behind the degradation are: 
(a) partial oxidation of the electrode, 
(b) formation of stable hydride phases,
(c) formation of the hydroxides surface layers (Mg(OH)$_2$ and La(OH)$_3$) hindering the diffusion of hydrogen into/or from the alloy, 
(d) expansion and contraction of 
the unit cell volumes during the hydrogenation and dehydrogenation processes leading to a pulverization of the alloy particles~\cite{miao_influences_2007, zhao_investigation_2008}.

It is well known that the cycle stability of the alloy is positively correlated with the capacity retention rate.
It is observed that this rate of the Gd free alloys is markedly lower than that of the alloy containing Gd or Co. 
For example, after 50 cycles, the capacity retaining rate ($R_{50}$) of the MA and annealed \lagdmgni{} alloys increase from 55\%, to 64\% and 95\% for $x = 0$, 0.25 and 0.5, respectively. 
In the case of the MA and annealed \lamghalfnico{} alloys the $R_{50}$ changes from 55\%, to 71\% and 65\% for $y = 0$, 0.5 and 1.5, respectively.
Phase composition is one of the most significant factors influencing the electrochemical performance of La-Mg-Ni-based alloys. For example, a high-Sm, Pr/Nd-free, and low-Co La$_{0.95}$Sm$_{0.66}$Mg$_{0.40}$Ni$_{6.25}$Al$_{0.42}$Co$_{0.32}$ alloy is reported as an \ab{}-type electrode for Ni-MH$_x$ batteries~\cite{cao_enhanced_2015}.
This induction melted alloy is composed of \lani{} (96.3\%) and LaNi$_5$ (3.7\%) phases, and shows high discharge capacity and excellent cyclic properties. 
The discharge capacity can reach 311.8~mAh/g and 227.0~mAh/g at large discharge currents of 2C and 5C, respectively, and 80\% of the capacity is retained after 239 cycles at 1C.
The effects of plasma milling (PM) and graphene addition in La$_{11.3}$Mg$_{6.0}$Sm$_{7.4}$Ni$_{61.0}$Co$_{7.2}$Al$_{7.1}$ alloy electrodes on the electrochemical properties and kinetics have been studied~\cite{ouyang_enhanced_2014}.
XRD analysis showed that the material was composed of Sm$_2$Co$_7$~\cite{cao_structural_2014}, LaNi$_5$ and a small amount of AlSm$_3$.
It has been found that the discharge capacity of this composite at  high discharge current density is significantly improved after the addition of graphene followed by PM.
At a rate of 5C, the discharge capacity of MH$_x$ electrode after PM for only 10~min has increased from 172.2~mAh/g to 243.9~mAh/g, and has further increased to 263.8~mAh/g after addition of graphene.
Moreover, the high-rate dischargeability and the exchange current density of the MH$_x$ electrodes has increased.
The recent progress on the application of dielectric barrier discharge plasma assisted milling in preparation of the energy storage materials has been reviewed by Ouyang~\textit{et al.}~\cite{ouyang_application_2017}.

The chemical substitution affects the kinetics of the hydrogen sorption reducing the same the time of the hydrogenation process~\cite{nowak_hydrogen_2018}.
Although the concentration and type of the considered substitutions do not affect substantially the maximum discharge capacities of our specimens, 
the alloying significantly improve the cyclic stability of the samples, see Figs.~\ref{fig:discharge} and \ref{fig:stability} and Table~\ref{tab:exp_data}. 
The best capacity retaining rates after the 50th cycle are obtained for the La$_{0.5}$Gd$_{1.0}$Mg$_{0.5}$Ni$_7$ and \lamghalfnicohalf{} electrodes. 
\subsection{XPS measurements}
The XPS technique provides information on the electronic structure of the valence band.
The valence band is affected by chemical composition and microstructure.
Furthermore, the valence band contains information about phase composition, local chemical disorder, impurities and local deformations.
\begin{figure}[ht]
\begin{center}
\includegraphics[trim = 0 0 0 0,clip,width=\columnwidth]{xps_exp.eps} 
\end{center}
\caption{\label{fig:xps_exp}
The experimental XPS valence band spectra (Al-K$_\alpha$) of bulk nanocrystalline \lani{}-based alloys: \lani{}, \lagdquartermgni{}, and  La$_{1.5}$Mg$_{0.5}$Ni$_{6.5}$Co$_{0.5}$.
All the intensities were normalized and separated by 0.3 arbitrary units for a better perception.  
}
\end{figure}
The bulk nanocrystalline \lagdquartermgni{} and \lamghalfnicohalf{} samples, 
identified as having the most promising electrochemical properties among the considered series,  
are investigated by the X-ray photoelectron spectroscopy (XPS).
The bulk nanocrystalline \lani{} sample is considered as the reference one.
The measured XPS valence band spectra of the considered alloys are presented in Fig.~\ref{fig:xps_exp}.
We observe that the full widths at half maxima (FWHM) of the valence bands of \lani{}, \lagdquartermgni{} and \lamghalfnicohalf{} are about 
2.8~eV, 2.5~eV and 2.8~eV, respectively.
In the XPS spectrum of the \lagdquartermgni{} we identify two main maxima, the first one around -0.9 eV and the second one around -1.9 eV.
Third much smaller maximum is observed at about -4.1 eV.
As the amount of Gd substitution in \lagdquartermgni{} is relatively low, the overall shape of its spectrum does not change much in respect to the \lamghalfnicohalf{} and \lani{} spectra.
The spectrum of the \lamghalfnicohalf{} is not much different from the reference spectrum of the \lani{} sample.
It is almost featureless and exhibits a single maximum at around -1.5~eV.
The main contributions to the all measured valence band spectra come from the 3$d$ electrons of Ni~\cite{werwinski_effect_2018}.
The details of the electronic valence bands will be discussed in the next section based on the results from the first-principles calculations.

According to semi-empirical models of metal hydride formation~\cite{bouten_heats_1980,griessen_heats_1988}, 
a justification for maximum hydrogen absorption capacity in metallic matrices has been proposed. 
These models show that the energy of the metal-hydrogen interaction depends both on geometric and electronic factors. 
One of the important electronic factor is the full width at half maximum (FWHM) of the valence band of the metallic host material.
In our case, the widths determined experimentally and theoretically for the considered samples are comparable, see Figs.~\ref{fig:xps_exp} and \ref{fig:xps_calc_comp}.  
Therefore, the maximum discharge capacities (C$_\mathrm{max}$) measured for these samples are practically the same. 
Furthermore, a small substitution of Ni by Co in \lamghalfnicohalf{} practically does not change the FWHM of valence band, but it improves the cycle stability, as shown in Figs.~\ref{fig:discharge} and \ref{fig:stability}.

\subsection{Density functional theory calculations}
In this section we continue the investigation of the electronic valence bands of two optimal compositions by using the DFT approach.
The details of our calculations are described in the methods section~\ref{ssec:dft_details}.
\begin{figure}[ht]
\centering
\includegraphics[trim = 0 0 0 0,clip,width=\columnwidth]{la_mg_gd_ni_co_xps_comparison.eps}
\caption{\label{fig:xps_calc_comp}
Calculated XPS spectra of \lani{}, \lamghalf{}, \lagdquartermgni{} and \lamghalfnicohalf{} phases (all in Ce$_2$Ni$_7$-type structures).
The electronic densities of states are calculated with scalar-relativistic PBE FPLO scheme treating disorder with the ordered compound method.
The $\delta$ parameter is set to 0.5.
}
\end{figure}
Fig.~\ref{fig:xps_calc_comp} shows the calculated spectra for the optimal compositions  \lagdquartermgni{} and \lamghalfnicohalf{} together with the results for the reference systems \lani{} and \lamghalf{}.
As the calculated XPS spectra are a combination of the contributions from all the valence band electrons, they can be resolved for types of orbitals, elements and sites.
\begin{figure}[ht]
\centering
\includegraphics[trim = 0 0 0 0,clip,width=\columnwidth]{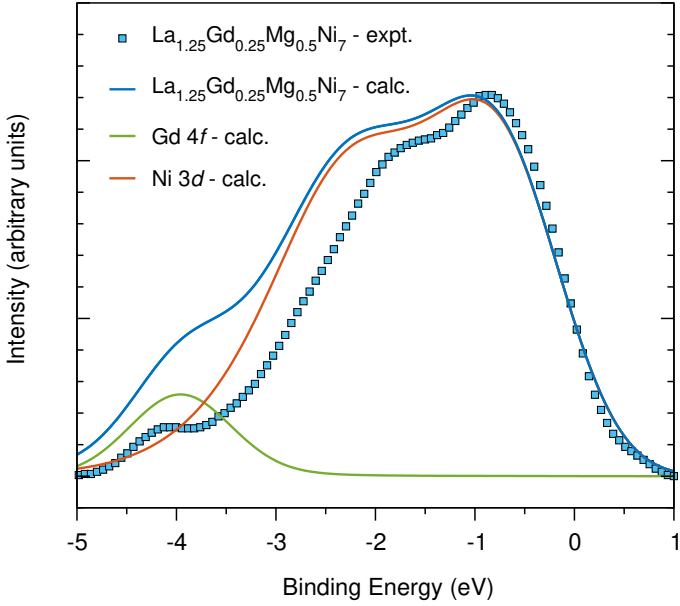}
\caption{\label{fig:xps_calc}
The calculated XPS spectra of \lagdquartermgni{} phase (in Ce$_2$Ni$_7$-type structure), together with contributions from Ni~3$d$ and Gd~4$f$ states.
The results are based on the spin-polarized scalar-relativistic PBE FPLO scheme treating chemical disorder with the ordered compound method.
The $\delta$ parameter is set to 0.5.
For comparison, the experimental XPS valence band is presented.
}
\end{figure}
Fig.~\ref{fig:xps_calc} presents the measured and calculated XPS spectra of \lagdquartermgni{}, together with the calculated contributions from Ni~3$d$ and  Gd~4$f$states.
The valence band of \lagdquartermgni{} is dominated by the Ni~3$d$ share, while the presence of Gd~4$f$ states around -4.0 eV explains the corresponding local maximum in the experimental spectra.
The Gd dopants contribute further to valence band with minor 5$d$ and 6$s$ shares, which are not shown.
\begin{figure}[h]
\begin{center}
\includegraphics[trim = 0 0 0 0,clip,width=1.0\columnwidth]{la_mg_gd_ni_co_DOS_comparison.eps} 
\end{center}
\caption{\label{fig:dos_comp}
Densities of states (DOS) for (a) \lani{}, (b) \lamghalf{}, (c) \lagdquartermgni{} and (d) \lamghalfnicohalf{} phases as calculated with scalar-relativistic PBE FPLO scheme treating disorder with the ordered compound method. 
Results for  (c) \lagdquartermgni{} show contributions from two spin channels (red and blue lines).
}
\end{figure}
The total DOSs for the considered compositions are presented in Fig.~\ref{fig:dos_comp}.
The advantage of the DOS plots over the XPS spectra is the lack of a broadening, which provides a more detailed picture of the electronic structure.
The DOSs of the all considered compositions reveal significant similarities.
A higher resolution of DOSs in respect to XPS spectra allows us to track changes in electronic structure even for small differences in concentration of Gd or Co dopants.
%
%----------------------------------van hove singularity----------------------------------
%
In all DOS plots from Fig.~\ref{fig:dos_comp} we observe two main maxima at about -1 eV and -2.5 eV, which are mainly related to the Ni~3$d$ orbitals.
Almost all systems, except the one with Co dopants, exhibit a characteristic narrow maximum close to the Fermi level, which can have a significant impact on the electrochemical properties of the considered alloys.
Similar peak has been observed for the bcc Ni~\cite{werwinski_effect_2018} and for some other Ni based compounds like for example MgC$_x$Ni$_3$~\cite{dugdale_electronic_2001} and YNi$_2$B$_2$X (X~=~B, C, N, and O)~\cite{lee_electronic_1994}.
Sharp narrow peak in DOS is called the van Hove singularity and happens to be located close to the Fermi level.
In opposition to the band-type states typical for valence band region,
the van Hove singularity can be understood as a fingerprint of an atomic-like state of electrons.
The van Hove singularity near the Fermi level observed in our calculation for La$_2$Ni$_7$ comes from the Ni 3$d$ states of most of the inequivalent atomic positions of Ni.
Unfortunately, the resolution of our XPS equipment is too low to investigate this narrow peak.
For example, similar features in valence bands have been observed previously with BIS technique for CeIr$_2$ and CeRu$_2$~\cite{allen_large_1983}.

%-------------------------------valence band shift----------------------------------
%
We conclude that small changes in the band structure in proximity of the Fermi level may have a significant effect on the electronic properties of the system.
The alloying of the system with a specific element which is increasing the occupation of the valence band should result in moving the Fermi level over the van Hove singularity.
In case of the considered La$_{1.5}$Mg$_{0.5}$Ni$_{7}$ composition, an increasing of the valence band occupation by alloying can be accomplished by substituting, for example, Ni by Cu.
The neutral copper atom has one more proton and one more electron than neutral Ni (atomic numbers: $Z_{Ni} = 28$ and $Z_{Cu} = 29$), 
thus Cu contributes to the La$_{1.5}$Mg$_{0.5}$Ni$_{7}$ system by increasing the occupation of the valence band.
Following this simple idea we perform the DFT calculations for alloying of Ni by Cu in the La$_{1.5}$Mg$_{0.5}$Ni$_{7-y}$Cu$_y$ system, where $y = 0.0, 0.5, 1.0, 1.5$.

\begin{figure}[h]
\begin{center}
\includegraphics[trim = 0 0 0 0,clip,width=1.0\columnwidth]{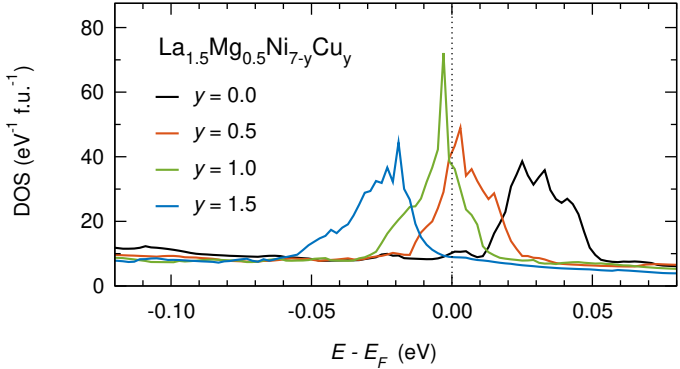} 
\end{center}
\caption{\label{fig:lamgnicu_dos_comp}
Effect of the composition on the electronic structure in the vicinity of Fermi level.
Densities of states (DOS) for La$_{1.5}$Mg$_{0.5}$Ni$_{7-y}$Cu$_y$ ($y = 0.0, 0.5, 1.0, 1.5$) phases as calculated with scalar-relativistic PBE FPLO scheme treating La/Mg disorder with the ordered compound method and Ni/Cu disorder with virtual crystal approximation (VCA).
}
\end{figure}
Previously, it has been shown that an a ferromagnetism in Y$_2$Ni$_7$ is due to a peak in DOS near the Fermi level~\cite{shimizu_electronic_1984}.
Fukase~\textit{et al.}, investigating the magnetic properties of \lani{} doped with Co or Cu, found that for more than 3\% of Cu in place Ni  the metamagnetic behavior disappears and ferromagnetic behavior appears and increase of Cu concentration up to 10\% induces a transition to the paramagnetic state~\cite{fukase_itinerant_2000}.
Tazuke~\textit{et al.} have later proposed that, if \lani{} has a similar peak as Y$_2$Ni$_7$ near the Fermi energy, the magnetic transitions observed along Cu alloying can be understood based on the evolution of electronic structure~\cite{tazuke_metamagnetic_2004}.  
Tazuke~\textit{et al.} have suggested that around 3\% of Cu concentration in \lani{} the Fermi energy reaches the maximum of the peak, satisfying the Stoner condition, and around 10\% Cu concentration the $E_{\mathrm{F}}$ goes beyond the peak, yielding the paramagnetic state.
In our notation (La$_{1.5}$Mg$_{0.5}$Ni$_{7-y}$Cu$_y$) 3\% and 10\% Cu concentrations would correspond to $y$ equal to 0.21 and 0.7, respectively.
Our results presented in Fig.~\ref{fig:lamgnicu_dos_comp} qualitatively confirm the explanation proposed by Tazuke~\textit{et al.}~\cite{tazuke_metamagnetic_2004}. 
In Fig.~\ref{fig:lamgnicu_dos_comp} we show that for Cu concentration $y \sim 0.5-1.0$ the position of the considered van Hove singularity falls on the Fermi level.
This observation may also explain the sensitivity of the electrochemical properties of \ab{}-type samples on small variations in chemical composition.
The real samples are, however, not as uniformed as we consider in theoretical models. 
Thus, we should expect more a qualitative agreement between the calculated and measured trends, rather than a quantitative agreement between the corresponding values.
The more advanced models should, among other, take into account the site preferences of the dopants and more accurately treat the chemical disorder, for example within the coherent potential approximation (CPA).

Here we present a CPA study of Co site preferences for La$_{1.5}$Mg$_{0.5}$Ni$_{6.5}$Co$_{0.5}$ in rhombohedral Gd$_2$Co$_7$-type system.
We observe the minimum energy for the Co dopants occupying the 9$e$ sites, but the 18$h$ sites are only slightly less energetically favored.
All of the five inequivalent Ni positions identified by Virkar and Raman are: 3$b$, 6$c_1$, 6$c_2$, 9$e$, 18$h$~\cite{virkar_crystal_1969,werwinski_effect_2018}.
The most preferred sites are at the same time the most capacious ones (9$e$, 18$h$).
\begin{figure}[h]
\begin{center}
\includegraphics[trim = 0 0 0 0,clip,width=1.0\columnwidth]{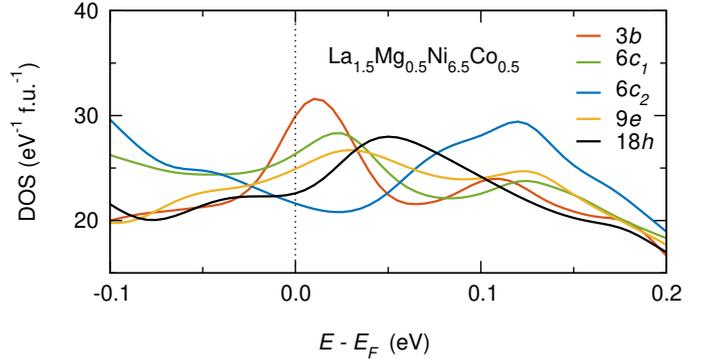} 
\end{center}
\caption{\label{fig:lamgnico_dos_comp}
Effect of Co site preference on the electronic structure in the vicinity of Fermi level.
Densities of states (DOS) for La$_{1.5}$Mg$_{0.5}$Ni$_{6.5}$Co$_{0.5}$ (Gd$_2$Co$_7$-type) phase as calculated with scalar-relativistic LDA FPLO scheme treating disorder with the coherent potential approximation (CPA).
}
\end{figure}
Fig.~\ref{fig:lamgnico_dos_comp} presents a series of DOSs in the vicinity of the Fermi level for Co dopants occupying five different Ni sites in the La$_{1.5}$Mg$_{0.5}$Ni$_{6.5}$Co$_{0.5}$ alloy.
The purpose of presenting these results is to show how different site occupations influence the DOS around the Fermi level.
The results from CPA confirm the limited application of our results from VCA and ordered structure approach to model the details of the electronic structure.
However, even the CPA models do not cover the microstructural features of the real samples.
Thus, they can serve more as a guide for performing conclusive experiments, rather to be a final settlement on the electronic properties of the bulk samples.
Fig.~\ref{fig:lamgnico_dos_comp} shows that the van Hove singularity is present for each of the considered site occupation.
The position of the narrow peak vary with the choice of the site.
For the case of the energetically most preferred 9$e$ site, the Fermi level is on the slope of the van Hove peak and close to its maximum.

\section{Summary and conclusions}

In this work we investigated the effect of substituting La by Gd and Ni by Co on electrochemical and electronic properties of the \lamghalf{} alloy. 
The experimental preparation and characterization were accompanied by the density functional theory (DFT) calculations. 
Two series of \lagdmgni{} ($x$~=~0.0, 0.25, 1.0) and \lamghalfnico{} ($y$~=~0.0, 0.5, 1.5) alloys were obtained by mechanical alloying. 
They were characterized by XRD and electrochemical measurements,
which showed that the obtained samples have a multi-phase character and consist mainly of the \ceni{}-type and \gdco{}-type structures of \lacommamgni{} phase with a minor contribution of LaNi$_5$ and La$_2$O$_3$ phases. 
The most promising electrochemical properties were identified for \lagdquartermgni{} and \lamghalfnicohalf{} alloy electrodes.
These two optimal compositions were a subject of the X-ray photoelectron spectroscopy (XPS) measurements.
The experimentally investigated XPS valence band spectra were interpreted based on the corresponding spectra calculated from first principles.
The theoretical analysis of valence bands focused on the van Hove singularity observed close to the Fermi level.
Presence of this narrow peak is suggested as one of the causes of variations in electrochemical properties of the systems with Gd and Co substitutions.

\section*{Acknowledgements}
Work supported by the National Science Centre Poland under the decision DEC-2014/15/B/ST8/00088. 
The computations were performed on the resources provided by the Poznań Supercomputing and Networking Center (PSNC).

\bibliography{lagdmgni7}   

\end{sloppypar}

\end{document}